\documentclass[trackchanges, twocolumn]{aastex701}

\begin{document}

\title{Gamma-Ray Modes in a Transitional Pulsar}

\correspondingauthor{Maksat Satybaldiev}
\email{maksat.satybaldiev@ntnu.no}

\author[0009-0001-3795-4049]{Maksat Satybaldiev}
\affiliation{Department of Physics, Norwegian University of Science and Technology, NO-7491 Trondheim, Norway}
\email{maksat.satybaldiev@ntnu.no}

\author[0000-0002-0237-1636]{Manuel Linares}
\affiliation{Department of Physics, Norwegian University of Science and Technology, NO-7491 Trondheim, Norway}
\affiliation{Departament de F{\'i}sica, EEBE, Universitat Polit{\`e}cnica de Catalunya, Av. Eduard Maristany 16, E-08019 Barcelona, Spain}
\email{manuel.linares@ntnu.no}
\begin{abstract}

Transitional millisecond pulsars (tMSPs) exhibit a unique subluminous disk state, at the crossroads between accretion and rotation power, where they switch between two distinct X-ray modes.
We present the discovery of gamma-ray modes in PSR J1023+0038, the first confirmed tMSP, from stacking Fermi-LAT data during the modes (which we identify using simultaneous X-ray observations).
Surprisingly, we find that gamma rays and X-rays are anticorrelated during this mode switching: the gamma-ray flux is higher in the X-ray low mode, and vice versa.
This contradicts the state-of-the-art model, which predicts bright gamma rays only in the X-ray high mode, from the interaction between the pulsar wind and surrounding disk via synchrotron and inverse Compton processes.
Because the pulsar wind is likely absent in the gamma-ray high (X-ray low) mode, which is also brighter in the radio band, we suggest that jet emission may be dominant in GeV gamma rays.
We also discuss an alternative interpretation where the gamma-ray high mode is powered by self-synchrotron Compton emission in the propeller state.

\end{abstract}

\keywords{\uat{Millisecond pulsars}{1062}, \uat{Neutron stars}{1108}, \uat{Gamma-rays}{637}, \uat{Low-mass X-ray binaries}{939}, pulsars: individual (PSR J1023+0038)}


\section{Introduction} \label{sec:introduction}

Transitional millisecond pulsars represent an evolutionary link between low-mass X-ray binaries (LMXBs) and rotation-powered millisecond pulsars (MSPs).
In the LMXB phase, the neutron star is spun up via angular momentum transfer from the accreted material \citep{binovatyi76, alpar82}.
When accretion stops, the fast rotating neutron star emits radio and gamma-ray millisecond pulsations. 
Transitional MSPs alternate between a radio pulsar and a disk state \citep{archibald09}. 
They exhibit a unique subluminous disk state with an average X-ray luminosity of $\sim10^{33}$~erg~s$^{-1}$, characterized by rapid switches between two modes and occasional flares \citep{linares14, bogdanov15, archibald15}.
The X-ray luminosity in this ``intermediate disk state" lies between that of neutron star LMXBs in the quiescent ($10^{30}-10^{32}$~erg~s$^{-1}$) and outburst states  \citep[$10^{35}-10^{38}$~erg~s$^{-1}$; see, e.g.,][]{wijnands15, heinke25}.
This provides a unique probe of underluminous accretion flows around neutron stars, at the crossroads between accretion- and rotation-powered phenomena.

Two groups of models have been proposed to explain the mode-switching behavior. 
In the first group, the high X-ray mode arises when the accretion disk enters the light cylinder, and the mode switches occur when the accretion disk moves outside the light cylinder \citep{linares14a, campana16, cotizelati18}.
In the other group of models, the high mode originates when the accretion disk is kept just outside the light cylinder \citep{papitto15, veledina19}.
The latter models are supported by the detection of optical and X-ray pulsations \citep{papitto19}, X-ray and optical polarization measurements \citep{baglio25}, and by recent numerical simulations of the pulsar-disk interaction
\citep{mignon-risse26,richard-romei26}.
However, the exact origin of the multiwavelength emission in the subluminous disk state is
still debated.

PSR J1023+0038 (J1023) was the first confirmed transitional MSP, discovered in 2007 when 1.7 ms radio pulsations were detected \citep{archibald09}.
Optical spectra taken in 2001 exhibited double-peaked emission lines \citep{thorstensen05,wang09}, suggesting that J1023 had an accretion disk and transitioned to a pulsar state between 2001 and 2007.
In 2013 June, radio pulsations from J1023 stopped being detected and an accretion disk was formed.
After the transition, broad double-peaked hydrogen emission lines appeared \citep{linares14ATel, stappers14}, 
and the X-ray and gamma-ray fluxes increased by factors of $\sim20$ and $\sim10$, respectively \citep{stappers14, takata14, linares14, torres17}.

The X-ray luminosity of J1023 switches stochastically between $(3.6\pm0.4)\times10^{32}$~erg~s$^{-1}$ in the low mode and $(2.40\pm0.04)\times10^{33}$~erg~s$^{-1}$ in the high mode \citep{bogdanov15, baglio23}.
X-ray, optical and UV pulsations were detected in the X-ray high and flaring modes \citep{archibald15, ambrosino17, papitto19, jaodand21, miravalzanon22}.
Simultaneous radio and X-ray observations revealed an anticorrelation between the X-ray and radio emission during mode switching \citep{bogdanov18, baglio23}.
In particular, drops in X-ray flux from the high to the low mode are accompanied by radio brightening over the duration of the low mode.
A similar radio and X-ray anticorrelation was found in the other candidate transitional MSP 3FGL J1544.6-1125 \citep{gusinskaia25}.
Radio emission of transitional MSPs can be explained by rapid ejections of plasma by the active pulsar from its magnetosphere \citep{bogdanov18}.
In an alternative scenario, the radio emission arises from a compact jet that is active in both modes, with discrete mass ejections superposed in the low mode \citep{baglio23}. 

The most recent models predict gamma-ray emission correlated with X-rays  during the mode switching \citep{veledina19}. 
Here, we study the GeV emission of the prototypical transitional MSP J1023, to investigate if it also shows gamma-ray modes.
We identify periods of low and high X-ray modes based on X-ray observations (Sec.~\ref{sec:xrayanalysis}) and then extract and analyze gamma-ray spectra corresponding to those modes (Sec.~\ref{sec:gammarays}).
We discover gamma-ray modes from J1023 anticorrelated with the X-ray modes (Sec.~\ref{sec:results}), and discuss our results in Section~\ref{sec:discussion}.

\begin{figure*}[ht!]
\includegraphics[width=1.0\textwidth]{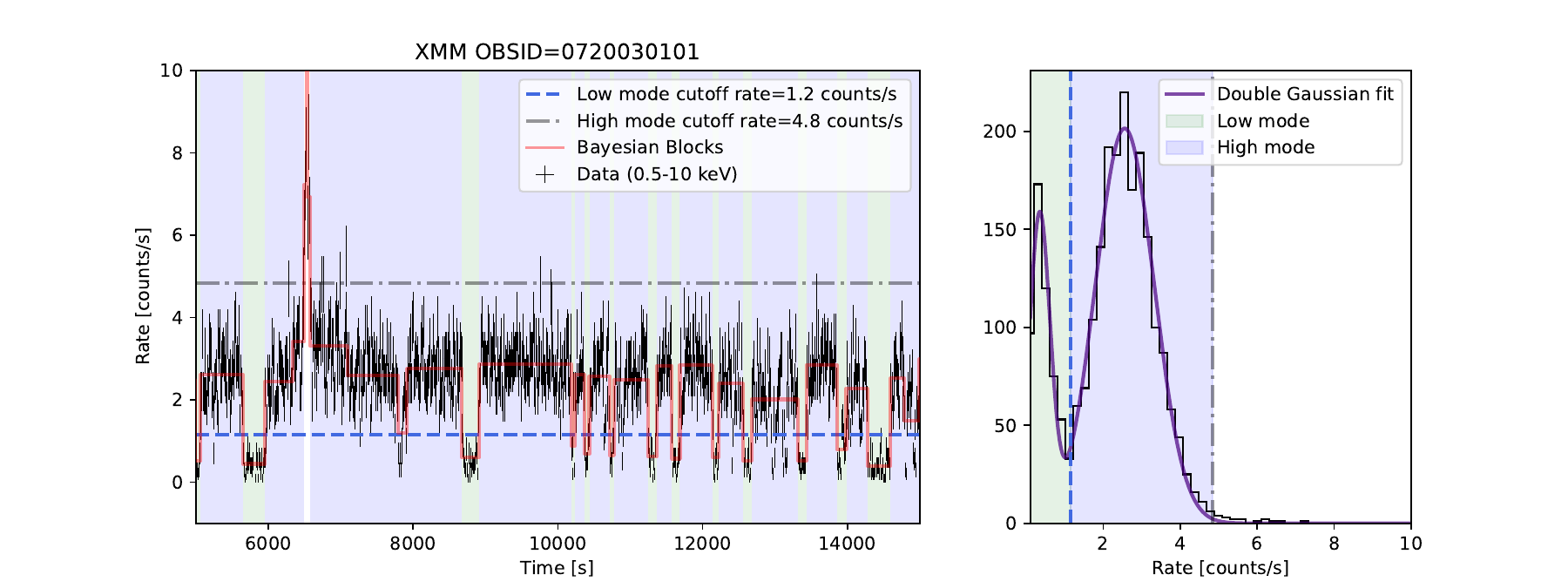}
\caption{Illustrative example of one XMM-Newton observation, where we identify high- and low-mode intervals.
Left: a segment of a light curve where J1023 switches between low and high modes, with a flaring episode. Black data points show  the EPIC-pn (0.5-10 keV) light curve and the red line is a Bayesian block representation of the light curve.  Right: count-rate histogram of the observation and double gaussian fit. Blue dashed  and gray dashed-dotted lines are the thresholds separating low-high and high-flaring states, respectively, defined in Section \ref{sec:xrayanalysis}. Green and blue regions show identified low and high mode time intervals.
\label{fig:rate_lc}}
\end{figure*}

\section{Observations and Data analysis} \label{sec:analysis}

The sensitivity of the Fermi Large Area Telescope (Fermi-LAT) does not allow individual observation of possible gamma-ray mode switching from J1023 on minute/hour timescales: 
LAT only detects about four photons per day in the disk state of J1023.
Instead, we can identify the modes based on simultaneous X-ray data, and accumulate gamma-ray photons separately for the two modes.
In this work, we apply this approach to study the gamma-ray behavior of J1023 during the disk state.

\subsection{X-Rays and mode selection} \label{sec:xrayanalysis}

We analyzed and extracted X-ray light curves from all available Neutron Star Interior Composition Explorer (NICER), X-ray Multi-Mirror Mission (XMM-Newton), Swift X-ray Telescope (Swift/XRT), and Nuclear Spectroscopic Telescope Array (NuSTAR) observations of J1023, taken after 2013 June when the system transitioned from the pulsar to the disk state.  
The light curves exhibit the expected mode-switching behavior, with occasional flares. 
To identify time intervals corresponding to the high and low modes, we proceeded as follows.

For each individual observation, we fitted the bimodal count-rate histograms with two Gaussian functions, obtaining means $(\mu_1,\mu_2)$ and standard deviations $(\sigma_1, \sigma_2)$ for X-ray low (1) and high (2) modes, respectively.
We defined the threshold separating low and high modes as $\mu_1+3\sigma_1$, and we used $\mu_2+3\sigma_2$ as the boundary between the high mode and flaring states. 

We then applied the Bayesian blocks algorithm \citep{scargle13}  to the NICER and XMM-Newton light curves (with a time resolution of 10 s) to select time intervals with a statistically significant flux difference (with p-value=0.05).
For each block, the count rate was computed as the average over the block duration.

Time blocks with an average count rate below $\mu_1+3\sigma_1$ were identified as X-ray low-mode intervals, while blocks with a count rate between $\mu_1+3\sigma_1$ and $\mu_2+3\sigma_2$ were classified as X-ray high-mode intervals.
This approach ensures that the identified intervals are robust against random count-rate fluctuations that may occasionally exceed the defined threshold values.
An example of a light curve with identified time intervals corresponding to the two modes is shown in Figure~\ref{fig:rate_lc}.

For the less sensitive NuSTAR and Swift/XRT data, where we used time bin sizes of 100s, the low- and high-mode intervals were identified directly from the light curves without applying Bayesian block segmentation, as explained below.
Across all instruments, J1023 spent approximately 22\% of the observed time in the low mode and 74\% in the high mode and showing flares in 4\% of the time (see Table~\ref{tab:observations} for details).

\begin{deluxetable}{llllccc}
\tablewidth{\textwidth}
\tabletypesize{\footnotesize}
\tablecaption{X-Ray observations and mode identification summary (see Sec.~\ref{sec:xrayanalysis} for details). The Nobs column gives the number of observations made with each telescope. The ``low mode" and ``high mode" columns show the number of identified time intervals, total exposure and fraction of the time J1023 was observed in that mode. The ``merged" row shows the number and total exposure time of time intervals combined across all X-ray instruments, accounting for possible overlaps. The ``Fermi livetime" row shows the corresponding cumulative exposure for those time intervals where J1023 was observed by Fermi-LAT (about 31\% of the merged exposure times, see Sec.~\ref{sec:searchLAT}).
\label{tab:observations}}
\tablehead{
Telescope & Years & Instrument (energy range) & Nobs & Total Exposure & X-Ray low mode & X-Ray high mode 
}
\startdata
NICER & 2017-2025 & XTI ($0.5-10$ keV) & 176\tablenotemark{a} & 1207 ks & 2030/277 ks/23\% & 4000/868 ks/72\% \\
XMM-Newton & 2013-2024 & EPIC-pn ($0.5-10$ keV) & 23\tablenotemark{b} & 816 ks & 877/193 ks/24\% & 1495/611 ks/75\% \\
Swift & 2013-2024 & XRT ($0.5-10$ keV) & 226\tablenotemark{c} & 340 ks & 263/65 ks/19\% & 359/263 ks/77\%  \\
NuSTAR & 2013-2021 & FPMA+B ($3-78$ keV) & 5\tablenotemark{d} & 313 ks & 242/53 ks/17\% & 384/242 ks/77\% \\ 
\hline
Merged & & & & & 3257/568 ks/22\% & 4818/1894 ks/74\% \\
Fermi livetime & & & & & 176 ks & 590 ks 
\enddata
\tablenotetext{a}{NICER target ID=1510, 6035, 6240, 6545, 7019.}
\tablenotetext{b}{XMM-Newton OBSID=0720030101, 0742610101, 0748390101, 0748390501, 0748390601, 0748390701, 0770581001, 0770581101, 0783330301, 0784700201, 0803620201, 0803620301, 0803620401, 0794580801, 0794580901, 0803620501, 0823750301, 0823750401, 0864010101, 0940220201, 0940220301, 0940220401, 0940220501.}
\tablenotetext{c}{Swift/XRT target ID=33012.}
\tablenotetext{d}{NuSTAR OBSID=30001027006, 30201005002, 80201028002, 90301006002, 30601005002.}
\end{deluxetable}

\subsubsection{NICER/XTI} \label{sec:nicer}
A total of 221 observations of J1023 were performed with the X-ray Timing Instument (XTI) on NICER \citep{nicer} between 2017 December and 2025 March. 
We analyzed these data using the standard \texttt{NICERDAS} v.15 software and \texttt{CALDB} v.20240206.
The raw, unfiltered data were processed using the \texttt{nicerl2} pipeline.
We extracted light curves in the $0.5-10$ keV energy band with \texttt{nicerl3-lc} and binned with $10$ s time resolution.
The \texttt{SCORPEON} model was used to estimate the background contribution.
We excluded 45 observations from our analysis, due to the low total exposure or high background that complicated the identification of X-ray modes.

From the 176 observations with a total exposure of 1.2 Ms, we identified 2030 low-mode and 4000 high-mode intervals, with accumulated exposures of 277 ks and 868 ks, respectively.

\subsubsection{XMM-Newton} \label{sec:xmm}
XMM-Newton \citep{xmm} observed J1023 25 times from 2013 October to 2024 June, after its transition to the disk state. 
In 23 of these observations, the EPIC-pn camera was active and operated in timing mode. 
We analyzed these observations using the Science Analysis Software (SAS v.22.1.0).
Background flaring intervals were identified from the $10-12$ keV light curves and excluded.
After selecting single-pixel events (\texttt{PATTERN==0}), we extracted $0.5-10$ keV light curves with a time resolution of 10 s from a box centered on the source (RAWX=38) with a width of 7 pixels.

From XMM-Newton observations with a total exposure of 816 ks, we identified 877 low-mode and 1495 high-mode intervals, corresponding to exposures of 193 ks and 611 ks.

\subsubsection{Swift/XRT} \label{sec:swift}

We used all 226 Swift/XRT \citep{swift, xrt} observations of J1023 with Swift target ID=33012, taken in photon counting (PC) mode between 2013 October and 2024 June while the system was in the disk state. We
did not include 22 observations taken with different target IDs during
the same time period, since the XRT pointing (and the effective count
rates) may differ. We reprocessed these using xrtpipeline (v.0.13.7,
within heasoft v.6.34), to create cleaned and calibrated event
files. The total resulting exposure was about 340 ks.

We extracted light curves in the full XRT (0.3-10~keV) energy band with a
bin size of 200 s, using a 30" radius region centered on
J1023. Exposure corrections, which account for dead pixels and bad
columns, can increase the effective count rates by typically 20-30\%
but do not change the overall count-rate distribution (see Section 2
of \citealt{linares14}). We estimate that the average background rate in the
same band is about 0.00099 counts~s$^{-1}$ (more than 5 times lower than the
lowest total count rates in the disk state). Thus, we use raw light
curves (not corrected for background or exposure maps) in order to
identify high- and low-mode periods in J1023.

We obtain an approximately bimodal XRT count-rate distribution in the disk state of J1023, consistent with that found from 2013 to 2014 data by \citet[][their Figure 1, left]{linares14}, but extending to 2013 to 2024 in our case. 
The same count-rate (or intensity) threshold of 0.1 counts~s$^{-1}$ 
separates the high and low modes effectively. With the extended
coverage, we are also able to separate flaring periods, which we define
as having count rates higher than 0.4 counts~s$^{-1}$. We then filter the
corresponding intensity ranges using xselect (v. 2.5b) to find the
times corresponding to low (0.005-0.1 counts~s$^{-1}$), high (0.1-0.4 counts~s$^{-1}$) and
flaring (0.4-1.0 counts~s$^{-1}$) modes.
The Swift/XRT data yielded 65 ks of low-mode and 263 ks of high-mode intervals.

\subsubsection{NuSTAR} \label{sec:nustar}

NuSTAR \citep{nustar} performed five observations of J1023 during its disk state from 2013 October to 2021 June, with a total exposure of 313 ks.
We analyzed and processed data using the \texttt{NUSTARDAS} v.2.1.5 package.
Raw event files were cleaned and calibrated with \texttt{nupipeline}. 
We extracted source and background light curves for the FPMA and FPMB modules separately in the $3-78$ keV energy band using \texttt{nuproducts} from 60" and 120" circular regions, respectively. 
The resulting FPMA and FPMB light curves were combined and binned at 100 s.

From the observations with a total exposure of 313 ks, we obtained 242 low-mode and 384 high-mode intervals, with exposures of 53 ks and 242 ks, respectively. 

In order to construct the broadband spectral energy distribution (SED; see Section~\ref{sec:sed}), we also extracted NuSTAR spectra of J1023 in its pulsar state. 
For that, we used three observations taken before the transition to the disk state (OBSID=30001027002, 30001027003 and 30001027005) with the total exposure of 95 ks.
The pulsar state spectra were obtained following the same procedure as for the light curves.

\subsection{Gamma Rays and Fermi-LAT} \label{sec:gammarays}

To characterize its long-term average GeV emission, we analyzed gamma-ray observations of J1023 performed by the LAT onboard the Fermi Gamma-ray Space Telescope mission \citep{atwood09}.
We performed binned likelihood analysis in the $0.1-100$ GeV energy range using the \texttt{fermipy} package \citep{wood17}.
We selected SOURCE class photons within a $15^\circ$-radius region of interest with zenith angles $\leq90^\circ$, and used \texttt{P8R3\_SOURCE\_V3} instrument response functions. 
The sky model was based on the 4FGL source catalog \citep{4FGL}, as well as the Galactic diffuse (\texttt{gll\_iem\_v07.fits}) and isotropic background (\texttt{iso\_P8R3\_SOURCE\_V3\_v1.txt}) models.

\subsubsection{Pulsar state}

We first analyzed LAT data collected from 2008 August to 2013 June (MJD 54682-56462), when J1023 was in the pulsar state.
We modeled the spectrum of J1023 using a simple power-law model,

\begin{equation}
    \frac{dN}{dE}=N_0\left(\frac{E}{E_0}\right)^{-\Gamma},
\end{equation} 

and a power law with an exponential cutoff,

\begin{equation}
    \frac{dN}{dE}=N_0\left(\frac{E}{E_0}\right)^{-\Gamma}\exp\left(-\frac{E}{E_c}\right),
\end{equation}

where $\Gamma$ is the photon index, $E_c$ is the cutoff energy and $E_0=1$ GeV.

Using the likelihood ratio test, we find that the additional cutoff component is not significant with $2\Delta\log\mathcal{L}=11.32$, in agreement with previous studies \citep{tam10, takata14, torres17}.
We find a photon index $\Gamma=2.42\pm0.09$ and a $0.1-100$ GeV energy flux $F^\gamma_\mathrm{PSR}=(5.9\pm0.7)\times10^{-12}$ erg cm$^{-2}$ s$^{-1}$.

\begin{deluxetable}{lcccc}
\tablewidth{\textwidth}
\tabletypesize{\small}
\tablecaption{Best-fit Gamma-Ray spectral parameters for J1023 in the different states and modes. Fluxes and luminosities are quoted in the $0.1-100$ GeV energy band. Luminosities use the distance of 1.37~kpc \citep{deller12}.
\label{tab:spectra}}
\tablehead{
State & Photon index $\Gamma$ & Cutoff energy $E_c$ (GeV) & Flux/$10^{-12}$ (erg cm$^{-2}$ s$^{-1}$) & Luminosity/$10^{33}$ (erg s$^{-1})$
}
\startdata
Pulsar state (PSR) & $2.42\pm0.09$ & ... & $5.9\pm0.7$ & $1.3\pm0.2$ \\
Disk state (DISK) & $2.02\pm0.03$ & $3.4\pm0.3$ & $48.4\pm0.7$ & $10.8\pm0.7$\\
X-ray low mode & $2.1\pm0.2$ & $3.4$ & $111\pm20$ & $25\pm5$\\
(D-XL) & $2.02$ & $1.9\pm1.1$ & $107\pm20$ & $24\pm5$\\
X-ray high mode & $2.8\pm0.4$ & $3.4$ & $29\pm7$ & $7\pm2$ \\
(D-XH) & $2.02$ & $0.8\pm0.5$ & $27\pm7$ & $6\pm2$ \\
\enddata
\end{deluxetable}


\begin{figure}[ht!]
\includegraphics[width=0.5\textwidth]{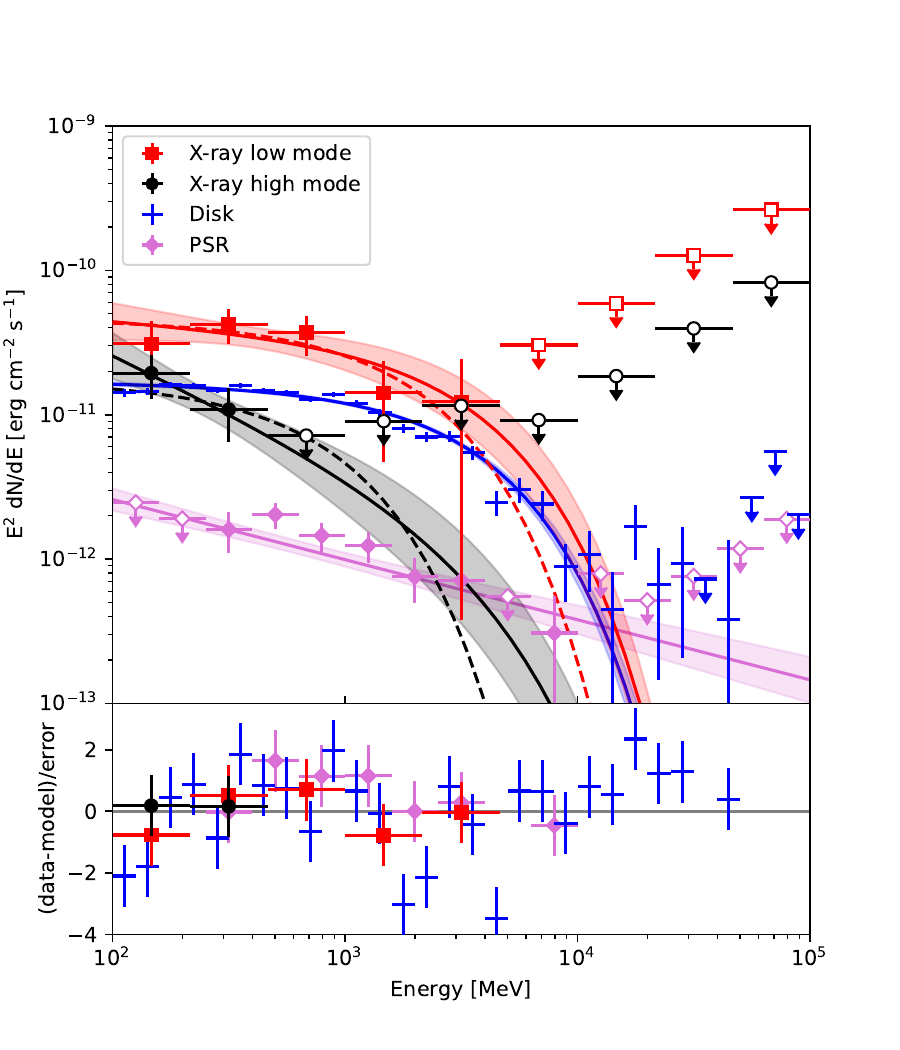}
\caption{Fermi-LAT spectral energy distributions of J1023 in the pulsar state (magenta diamonds), average disk state (blue crosses), X-ray low mode (red squares), and X-ray high mode (black circles). Solid symbols show data points with TS$>5$ and corresponding $1\sigma$ uncertainties, and open symbols show $95\%$ confidence level upper limits. Solid lines and shaded regions show best-fit models with the cutoff energy $E_c$ fixed at the average disk state value and their $1\sigma$ uncertainties, and dashed lines show best-fit models with the photon index $\Gamma$ fixed at the average disk state value.
\label{fig:seds}}
\end{figure}

\subsubsection{Disk state}
\label{sec:disk_state}

For the disk state, we analyzed LAT data collected from 2013 June to 2025 November (MJD 56462-60991).
We find that a spectral fit with an exponentially cutoff power law
is strongly preferred 
over a simple power law, with $2\Delta\log\mathcal{L}=186.5$, indicating that the cutoff is highly significant (consistent with \citealt{takata14, torres17}).
We obtain $\Gamma=2.02\pm0.03$ and 
$E_c=3.4\pm0.3$~GeV with a total energy flux of $F^\gamma_\mathrm{DISK}=(48.4\pm0.7)\times10^{-12}$ erg cm$^{-2}$ s$^{-1}$, corresponding to an $\sim8$-fold increase compared to the pulsar state.

The average SEDs of J1023 in the pulsar (magenta) and disk (blue) states are shown in Figure~\ref{fig:seds}.

\begin{figure*}[ht]
\gridline{
  \fig{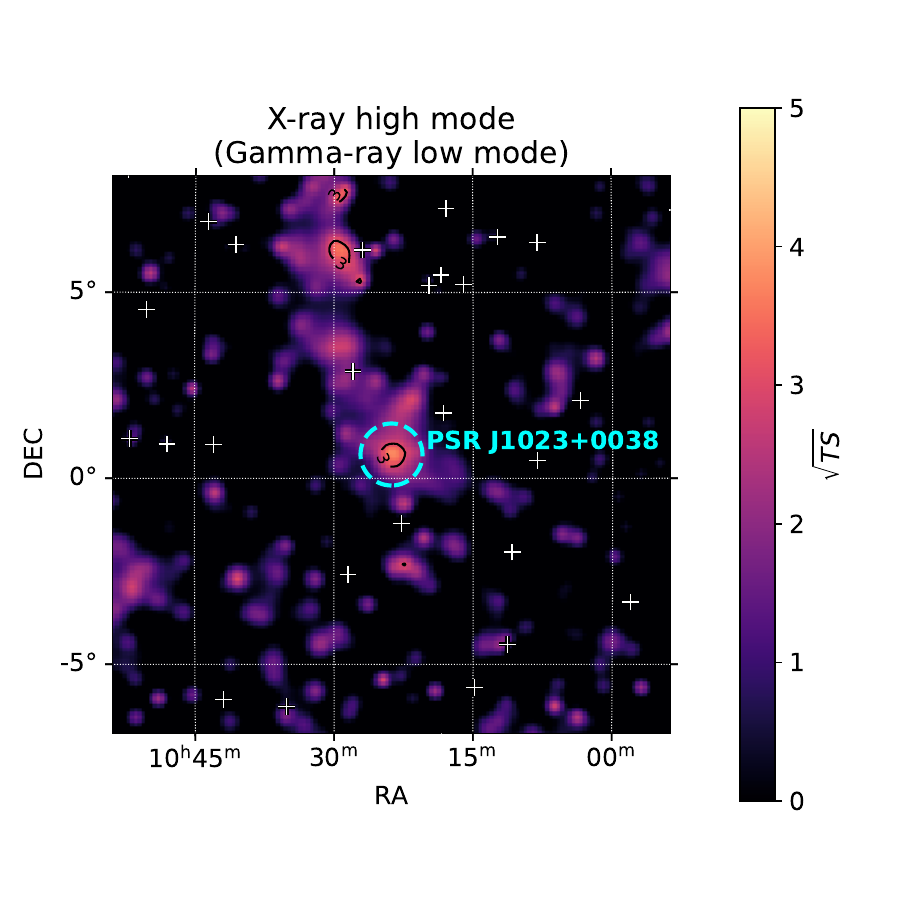}{0.5\textwidth}{}
  \fig{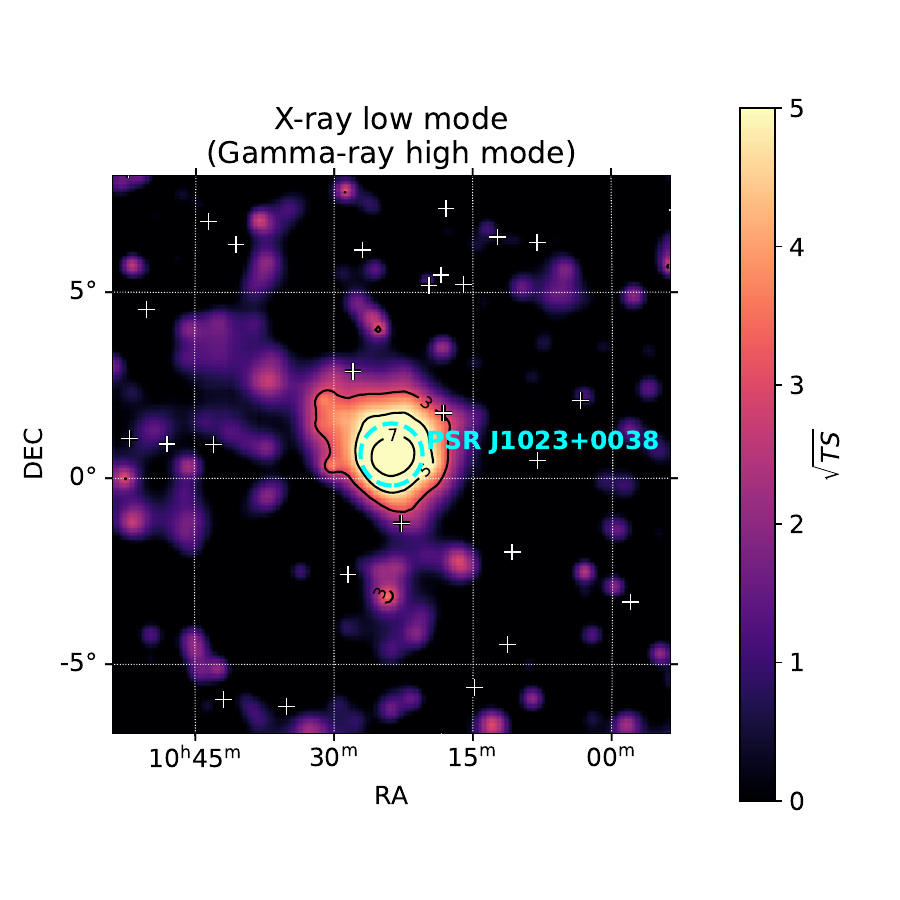}{0.5\textwidth}{}
}
\caption{Test statistics maps of the region of interest with J1023 removed from the models for X-ray high mode (left) and X-ray low mode (right). The map is produced by adding a test source modeled with a simple power-law spectral model with a photon index fixed at $\Gamma=2$. Dashed cyan circles show the location of J1023. White crosses correspond to the locations of the modeled sources.}
\label{fig:TS_maps}
\end{figure*}

\subsubsection{Search for Gamma-Ray modes}
\label{sec:searchLAT}

In order to search for gamma-ray modes, we converted the high- and low-mode time intervals identified from X-ray observations, from the corresponding mission elapsed times (METs) to Fermi-LAT METs, without applying barycentric corrections\footnote{The expected light-travel time differences between the spacecrafts are $\lesssim1$ s, which is smaller than the minimum light-curve time bin size of $10$ s. Thus, barycentric corrections are not needed for this analysis.}.
Finally, we merged the resulting time intervals, accounting for overlaps (in a few cases, X-ray observations were performed simultaneously).
In total, we obtained 3257 low-mode and 4818 high-mode intervals with merged exposures of 568 ks and 1894 ks, respectively. 
A full summary of identified intervals and corresponding exposures is given in Table~\ref{tab:observations}.

After identifying and merging high- and low-mode intervals, we selected Fermi-LAT events and generated livetime cubes for each time interval.
We then combined all high and low mode livetime cubes using \texttt{gtltsum} and performed a binned likelihood analysis for the merged high- and low-mode intervals.
By  excluding times when J1023 was occulted by the Earth (zenith angle $<90^\circ$) and integrating the livetime cubes over the inclination angle (the angle between the direction to a source and the LAT normal) from $0^\circ$ to $90^\circ$, we find that Fermi-LAT observed J1023 for 176 ks in time intervals that we can identify as the low mode, and 590 ks in times identified with the high mode.

\section{Results} \label{sec:results}

We detect GeV emission from J1023 in both X-ray modes, and we find that the gamma-ray flux differs between them.
Thus, we discover gamma-ray modes in a transitional MSP.
Interestingly, we  find the gamma-ray modes to be anticorrelated with X-rays: the gamma-ray flux is lower in the X-ray high mode, and vice versa.
Figure~\ref{fig:TS_maps} shows the Test Statistic (TS) maps for the two mode interval selections, assuming a test source with a photon index of $\Gamma=2$, where J1023 has been removed from the model.
Despite the shorter exposure during the X-ray low mode, J1023 is detected more significantly in this mode with TS$=81.4$, compared to TS$=24.8$ in the high X-ray mode. 
Thus, we conclude that J1023 is brighter in (GeV) gamma rays when it is fainter in (keV) X-rays.

In the discussion that follows, we refer to the pulsar and disk states with the subscripts ``PSR" and ``DISK", respectively.
We refer to the X-ray low and high modes as ``D-XL" and ``D-XH", respectively (corresponding to the gamma-ray high and low modes). 
We also use superscripts to indicate the band where fluxes/luminosities are measured ($\gamma$ and $X$). 

Table~\ref{tab:spectra} shows the best-fit spectral parameters for both modes.
Due to the limited statistics in the LAT spectra of both modes, we fixed the cutoff energy $E_c$ to the value obtained for the disk state (3.4~GeV) when fitting the individual mode SEDs.
In the X-ray low mode (gamma-ray high mode), the spectrum is best modeled with a photon index of $\Gamma=2.1\pm0.2$ and a $0.1-100$ GeV flux of $F^\gamma_\mathrm{D-XL}=(111\pm20)\times10^{-12}$ erg cm$^{-2}$ s$^{-1}$, with 71 predicted counts from J1023. 
Compared to the average disk state, the source is twice as bright in the LAT band, with no significant change in the spectral slope.

In the X-ray high mode (gamma-ray low mode), the spectrum is best modeled with a photon index of $\Gamma=2.8\pm0.4$ and a $0.1-100$ GeV flux of $F^\gamma_\mathrm{D-XH}=(29\pm7)\times10^{-12}$ erg cm$^{-2}$ s$^{-1}$. 
Despite the longer exposure time accumulated in the X-ray high mode, the number of predicted LAT counts is similar (75).
The LAT spectrum is slightly softer in the gamma-ray low mode, and the 0.1-100 GeV flux is fainter by a factor of 3.8 and 1.7 compared to the X-ray low mode and the average disk state, respectively.
Additionally, in the X-ray high mode the source is a factor of 5 brighter compared to the pulsar state.

To quantify the change in the spectral cutoff energy, we performed alternative fits in which the photon index was fixed to the average disk state value and the cutoff energy was left free to vary.
Both parameterizations yield statistically equivalent fits with no significant preference for either model.
We find $E_c=1.9\pm1.1$ GeV  in the X-ray low mode and $E_c=0.8\pm0.5$ GeV in the X-ray high mode.
In both modes,
the cutoff energy is not well constrained, but it seems to occur at lower energies than the average disk state spectrum ($E_c=3.4\pm0.3$ GeV; see Section~\ref{sec:disk_state}).

The SEDs of J1023 in the low and high modes are shown in Figure~\ref{fig:seds}.
For each energy bin with TS$>5$, we show data points with $1\sigma$ uncertainty error bars; when TS$<5$, we plot $95\%$ confidence level upper limits.
For the two modes, we show the best-fit power law with an exponential cutoff models with either the cutoff energy $E_c$ (solid lines) or the photon index $\Gamma$ (dashed lines) fixed to the average disk state values.
From this, we see that the gamma-ray emission in both modes is brighter than that in the pulsar state.
As expected, the average disk state SED lies between the SEDs of the two modes.

The flux difference between the two gamma-ray modes is $F^\gamma_\mathrm{D-XL}-F^\gamma_\mathrm{D-XH}=(82\pm21)\times10^{-12}$ erg cm$^{-2}$ s$^{-1}$, significant at the $3.9\sigma$ level.
This shows that the gamma-ray and X-ray fluxes are anticorrelated during the disk state of J1023.

\begin{figure*}[ht!]
\includegraphics[width=1.0\textwidth]{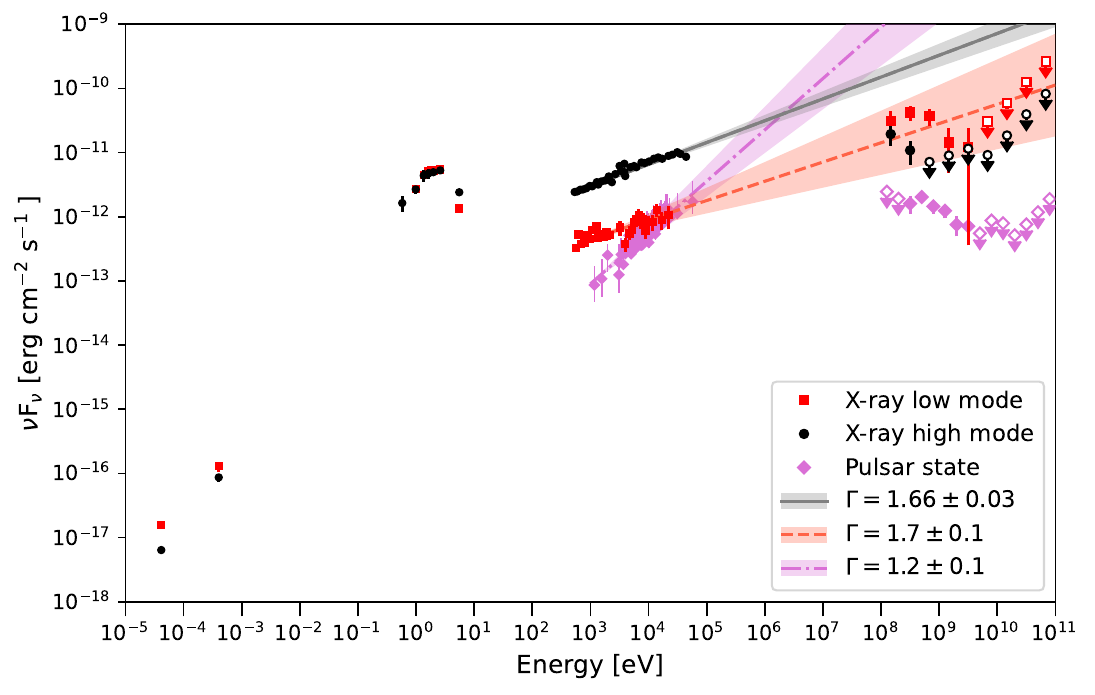}
\caption{The broadband SED of J1023. Black points show X-ray high-mode data, red point X-ray low-mode data and magenta points show the pulsar state data.
The lines and shaded areas show power-law fits to the X-ray data and corresponding 1$\sigma$ uncertainties on the photon index. 
\label{fig:nufnu}}
\end{figure*}

\citealt{baglio19} observed a near-infrared flare about 10~s after one 
X-ray low-to-high-mode transition. To verify that the enhanced gamma-ray emission does not arise only around the mode-switching intervals, we reanalyzed the Fermi-LAT data after trimming 10 s from the beginning and end of each high and low mode interval, to exclude these transitions. We find consistent fluxes and photon indices within $1\sigma$, confirming that the gamma-ray emission is associated with the full mode intervals and not only with the mode-switching transitions.

\section{Discussion} \label{sec:discussion}
\subsection{Energetics} \label{sec:energetics}

Assuming a distance of  $d=1.37\pm0.04$ kpc \citep{deller12}, we calculate the gamma-ray luminosities of J1023 in its different states and modes in the $0.1-100$ GeV energy range, and list them in Table~\ref{tab:spectra}.
The gamma-ray luminosities in both modes exceed the pulsar state value, showing that even if rotation-powered gamma-ray emission is present in either mode, additional emission components are required. 
For comparison, the spin-down luminosity of J1023 is $\dot{E}=4.43\times10^{34}$ erg s$^{-1}$ \citep{archibald13}.
The gamma-ray luminosity of J1023 in the X-ray low mode, when the disk is thought to enter the light cylinder, is 56\% of $\dot{E}$. 
This efficiency is higher than that in the  pulsar state of J1023 (3\% of $\dot{E}$) and than those observed in most MSPs, suggesting that
an extra energy source is needed 
to power the gamma-ray emission in this mode (beyond rotation power; see Section~\ref{sec:gammahigh}).

In the X-ray high mode, the gamma-ray luminosity is 16\% of $\dot{E}$.
Comparing with the typical gamma-ray efficiency of MSPs of $\sim10\%$ \citep{3PC}, the X-ray high-mode gamma-ray emission is roughly consistent with the rotation-powered emission of nontransitional MSPs.
In this case, if the pulsar wind hits the inner accretion disk and produces pulsed GeV emission \citep{veledina19}, we may expect gamma-ray pulsations to be detectable in the X-ray high mode. 

The X-ray low- to X-ray high-mode flux ratio in gamma rays is $F^\gamma_\mathrm{D-XL}/F^\gamma_\mathrm{D-XH}=3.8\pm1.2$, and the X-ray high- to X-ray low-mode flux ratio in X-rays is $F^X_\mathrm{D-XH}/F^X_\mathrm{D-XL}=6.6\pm0.6$,
indicating a smaller change in gamma-ray flux compared to that seen in X-rays.

The $0.5-10$ keV X-ray luminosity in the pulsar state is $L^X_\mathrm{PSR}=(9.4\pm0.4)\times10^{31}$~erg s$^{-1}$ \citep{archibald10}, and the $0.3-10$ keV X-ray luminosities in the X-ray low and high modes are $L^X_\mathrm{D-XL}=(3.6\pm0.4)\times10^{32}$~erg s$^{-1}$ and  $L^X_\mathrm{D-XH}=(2.40\pm0.04)\times10^{33}$~erg s$^{-1}$, respectively  \citep{baglio23}.
Thus, while the gamma-ray to X-ray flux (luminosity) ratio is $\log(F^\gamma_\mathrm{PSR}/F^X_\mathrm{PSR})=1.14\pm0.07$ in the pulsar state, we find $\log(F^\gamma_\mathrm{D-XH}/F^X_\mathrm{D-XH})=0.4\pm0.1$ in the X-ray high mode and $\log(F^\gamma_\mathrm{D-XL}/F^X_\mathrm{D-XL})=1.8\pm0.1$ in the X-ray low mode.
For the population of redback systems, the flux ratio is $\log(F^\gamma_\mathrm{RB}/F^X_\mathrm{RB})=1.6\pm0.8$ \citep{koljonen25},
similar to that of J1023 in the pulsar state and in the low X-ray mode.
We conclude that the pulsar and disk states of J1023 cannot be distinguished based on the gamma-ray to X-ray flux ratio alone.

 \subsection{Broadband SED} \label{sec:sed}

In the disk state, we have found and reported an {\it anticorrelation} between the X-ray and gamma-ray fluxes.
Because the X-ray and radio fluxes were found to be anticorrelated \citep{bogdanov18}, we infer a {\it positive} correlation between the gamma-ray and radio luminosities of the modes.
To construct the broadband SED of J1023, we used the radio to X-ray data in both modes from \citealt{baglio23}, the pulsar state Swift/XRT spectrum from \citealt{linares14}, and the NuSTAR pulsar state spectrum (Section~\ref{sec:nustar}) and the gamma-ray data from this work (Section~\ref{sec:gammarays}).
Figure~\ref{fig:nufnu} shows the broadband SED of J1023 in its two X-ray modes and in the pulsar state. 
Lines show the power-law fits to the X-ray data. 
We used the disk state fits with $\Gamma=1.66\pm0.03$ in the high X-ray mode and $\Gamma=1.7\pm0.1$ in the low X-ray mode \citep{baglio23}, and $\Gamma=1.2\pm0.1$ in the pulsar state \citep{li14}.

The X-ray and gamma-ray emission in the pulsar state have different origins. 
The X-ray emission originates from synchrotron emission of electrons and positrons accelerated in the intrabinary shock \citep[IBS;][]{romani16, wadiasingh17, wadiasingh18, vandermerwe20}, while the gamma-ray emission originates in the current sheet just outside the light cylinder \citep{philippov22, kalapotharakos23, hakobyan23, 3PC, cerutti25, kalapotharakos26}.
Thus, it is not surprising that the extrapolation of the X-ray power law lies well above the gamma-ray (LAT) data.
We expect a spectral cutoff (or synchrotron cooling break) of the IBS emission between the observed X-ray and gamma-ray energies, somewhere in the $\sim$100~keV-100~MeV range \citep{sullivan25}.

While the extrapolation of the low X-ray mode power-law fit can explain the gamma-ray data in this mode, the high X-ray mode fit overestimates the gamma-ray emission, suggesting again a spectral cutoff between  $\sim100$ keV and $\sim100$ MeV.

\subsection{Low Gamma-Ray  mode emission} \label{sec:gammalow}
\citealt{veledina19} proposed that the gamma-ray emission of J1023 comes entirely from the X-ray high mode. The synchrotron radiation of pulsar wind particles heated by collision with the disk, as well as synchrotron and inverse Compton emission of the disk plasma heated by the pulsar wind, may produce pulsed GeV emission. 
Our finding that the gamma-ray emission is brighter in the X-ray low (gamma-ray high) mode rules out this scenario.

Synchrotron and Compton emission could still 
be produced in the X-ray high (gamma-ray low) mode, at the inner disk heated by the pulsar wind just outside the light cylinder.
This could yield the gamma-ray luminosity that we measure ($L^\gamma_\mathrm{D-XH}=(7\pm2)\times10^{33}$~erg s$^{-1}$, see Table~\ref{tab:spectra}): about 5 times higher than that of the pulsar state.
The detection of gamma-ray pulsations in this gamma-ray low mode would indicate that the emission region corotates with the neutron star.
%
However, our results show that an additional gamma-ray emission mechanism 
is needed in the X-ray low (gamma-ray high) mode.




\subsection{High Gamma-Ray mode emission} \label{sec:gammahigh}
Here we consider possible origins for the gamma-ray emission in the brighter gamma-ray high mode.
A full, quantitative model of the gamma-ray modes that we have discovered is beyond the scope of this work.

\subsubsection{Truncated disk} \label{sec:pushed_disk}

\citealt{papitto19} propose that the disk is pushed outside and far from the light cylinder in the X-ray low (gamma-ray high) mode. In this case, the gamma-ray emission would come predominantly from the rotation-powered pulsar.
In Section~\ref{sec:energetics} we showed that an extra energy source is needed to power gamma rays in the gamma-ray high mode, which disfavors this scenario.
The gamma-ray luminosity in the X-ray low mode is about 20 times higher than that of the pulsar state (Table~\ref{tab:spectra}); this seems very difficult to sustain if the disk is truncated far out.

\subsubsection{Propeller} \label{sec:propeller}

In the model proposed by \citealt{veledina19}, the X-ray low (gamma-ray high) mode occurs when the disk moves inside the light cylinder and the system enters the so-called propeller state. 
In this regime, electrons accelerated in the shocked environment at the boundary between the disk and magnetosphere can produce gamma rays via self-synchrotron Compton (SSC) emission \citep{papitto14, papitto15}. The propelled outflowing material can also produce synchrotron radiation at radio frequencies, explaining the correlated radio and gamma-ray emission. 
%
Here we note that this same SSC mechanism could power the gamma-ray emission in the X-ray low (gamma-ray high) mode, with $L^\gamma_\mathrm{D-XL}=(25\pm5)\times10^{33}$~erg s$^{-1}$.

\subsubsection{Jet emission} \label{sec:jet}

\citealt{baglio23} argue that a continuously active compact jet powers the X-ray and radio emission in both modes.
In the high mode, the jet's X-ray emission contribution to the X-ray emission is small ($\sim8-18\%$), with most of the flux arising from the shocked inner accretion disk region \citep{veledina19, papitto19, baglio23}.
In the low X-ray mode, the X-ray emission arises almost entirely from the jet, while the radio emission has contributions from the jet and  discrete matter ejections.

If the jet power-law emission extends to gamma-ray energies, it may explain the gamma-ray emission in the X-ray low mode (Figure~\ref{fig:nufnu}). 
However, since the jet is present in both modes, the X-ray high mode should also have the same gamma-ray jet contribution. 
But the observed gamma-ray flux in this mode is lower.
This can be interpreted in several ways.
(1) The jet emission explains the X-ray emission but does not reach gamma-ray energies, having a cooling break between $\sim100$ keV and $\sim100$ MeV.
In this case, some other emission process must be responsible for the observed gamma-ray modes.
(2) The jet produces gamma-ray emission only in the X-ray low mode,  either because of a spectral break between $\sim100$ keV and $\sim100$ MeV present only in the X-ray high mode, or because the jet is not active in that mode.
(3) The X-ray low-mode emission is not produced by the jet.
Indeed, to explain the X-ray low-mode emission by the compact jet, \citealt{baglio23} require the jet break frequency to be $\nu_{br}=(2.4^{+1.9}_{-1.0})\times10^{13}$ Hz. 
\citealt{koljonen25} show that the jet break frequency of J1023 is unconstrained, with a potential range between $\sim10^{11}$ Hz and $\sim10^{13}$ Hz.
If the break occurs at lower frequencies, the compact jet would not explain the observed X-ray and gamma-ray emission, and some other emission processes would be responsible for the X-ray and gamma-ray modes.

If the gamma-ray emission in the X-ray low mode is produced by the jet, the spectral cutoff observed in the LAT band at $1.9\pm1.1$ GeV ($4.6\times10^{23}$~Hz) could be interpreted as the synchrotron cooling break, which has a frequency given by
\begin{equation}
    \nu_{cool}=\frac{3\gamma^2_{cool}eB_{cool}}{4\pi m_ec},
    \label{eq:nu_cool}
\end{equation}
where $\gamma_{cool}$ is the Lorentz factor of the cooling electrons, $m_e$ and $e$ are the mass and charge of an electron, $c$ is the speed of light, and $B_{cool}$ is the magnetic field in the cooling zone \citep{rybicky79}.
The synchrotron cooling timescale is 
\begin{equation}
\tau=\frac{6\pi m_e c}{\sigma_TB_{cool}^2\gamma_{cool}},
\label{eq:tau_cool}
\end{equation}
where $\sigma_T$ is the Thompson cross section.
Assuming $\tau\lesssim10$ s \citep[the existing limits on the X-ray high-to-low-mode transition time;][]{baglio23} and solving Equations \ref{eq:nu_cool} and \ref{eq:tau_cool}, we can place limits on the magnetic field 
\begin{equation}
\small
    B_{cool}\gtrsim0.4\left(\frac{10\text{ s}}{\tau}\right)^{2/3}\left(\frac{4.6\times10^{23}\text{ Hz}}{\nu_{cool}}\right)^{1/3}\text{ G},
\end{equation}
and a Lorentz factor
\begin{equation}
\small
    \gamma_{cool}\lesssim5\times10^8\left(\frac{\nu_{cool}}{4.6\times10^{23}\text{ Hz}}\right)^{1/2}\left(\frac{0.4\text{ G}}{B_{cool}}\right)^{1/2}.
\end{equation}
From the light-travel distance, we can also infer that the cooling zone is located at $z_{cool}\lesssim3\times10^{11}$~cm from the neutron star.

To estimate the magnetic field and size of the jet base, we use the optical thick-to-thin break values from \citealt{baglio23}:
the break frequency $\nu_{br}=2.4\times10^{13}$~Hz, the optically thin spectral index $\alpha_{thin}=-0.78$, and the flux density at the break frequency of $F_{br}=0.257$~mJy.
The value of $\alpha_{thin}$ corresponds to a nonthermal electron power-law index of $\sim2.5$.
Following Equations (3) and (6) from \citealt{koljonen25}, we estimate the jet base size and magnetic field to be $R_0\approx4\times10^7$~cm and $B_0\approx4\times10^{4}$~G.
For a jet opening angle $\phi<30^\circ$, the jet base is located at $z_0\gtrsim1.5\times10^8$~cm.
For $B_{cool}=0.4$~G and a simple steady conical jet geometry \citep{blandford79}, in which the magnetic field scales with distance from the jet base as $B\propto z^{-1}$, the cooling zone would be located well beyond the light-travel distance. 
This suggests a more complex nonsteady jet geometry and magnetic field structure in J1023.

\section{Conclusions} \label{sec:conclusions}

In this work, we analyzed all available NICER, XMM-Newton, Swift/XRT, and NuSTAR X-ray data, identifying time intervals corresponding to the high and low modes observed in the subluminous disk state of J1023. 
Using these intervals, we performed a Fermi-LAT analysis and detected significant gamma-ray emission in both modes.
We find that the gamma-ray fluxes differ between the two modes and are anticorrelated with those in X-rays. This represents the first detection of gamma-ray modes in a transitional MSP.

J1023 is brighter in gamma rays in the low X-ray mode (where it is detected up to 4 GeV) than in the X-ray high mode (where it is  detected up to 0.5 GeV).
Because our results imply a correlation between gamma-ray and radio luminosity, we discuss a simple jet model for the X-ray low mode.
Alternatively, the gamma-ray modes could be driven by transitions to the propeller state, or a different physical mechanism.

\begin{acknowledgments}
M.S. is grateful to Karri Koljonen for sharing the broadband SED data, to Egor Podlesnyi for the help with combining multiepoch Fermi-LAT data, and to Raphaël Mignon-Risse for useful discussions. We thank the anonymous referee for constructive comments, which helped improve this manuscript.

This project has received funding from the European Research Council (ERC) under the European Union’s Horizon 2020 research and innovation programme (grant agreement No. 101002352, PI: M. Linares).

This research has made use of data and software provided by the High Energy Astrophysics Science Archive Research Center (HEASARC), which is a service of the Astrophysics Science Division at NASA/GSFC.

This work made use of Claude (Anthropic) for writing and debugging code.
\end{acknowledgments}




%
\facilities{Fermi (LAT), NICER, XMM, Swift (XRT), NuSTAR}

\software{Astropy \citep{2013A&A...558A..33A,2018AJ....156..123A,2022ApJ...935..167A}, Fermi Science Tools \citep{2019ascl.soft05011F}, Fermipy \citep{wood17}, NumPy \citep{harris2020array}, SciPy \citep{2020SciPy-NMeth}, Matplotlib \citep{Hunter:2007}, HEASOFT, SAS}



\bibliographystyle{aasjournalv7}



\end{document}